# Critical Temperature for the Nuclear Liquid-Gas Phase Transition (from Multifragmentation and Fission)

V. A. Karnaukhov,[1,*] H. Oeschler,[2] A. Budzanowski,[3] S. P. Avdeyev,[1] A. S. Botvina,[4]

E. A. Cherepanov,[1] W. Karcz,[3] V. V. Kirakosyan,[1] P.A. Rukoyatkin,[1]

I. Skwirczyńska,[3] E. Norbeck[5]

[1] Joint Institute for Nuclear Research, Dubna, Russia;

[2] Institut für Kernphysik, Darmstadt University of Technology, Darmstadt, Germany;

[3] H. Niewodniczanski Institute of Nuclear Physics, Cracow, Poland;

[4] Institute for Nuclear Research, Moscow, Russia;

[5] University of Iowa, Iowa City, Iowa, USA

Critical temperature $T_c$ for the nuclear liquid-gas phase transition is estimated both from the multifragmentation and fission data. In the first case, the critical temperature is obtained by analysis of the IMF yields in $p$(8.1 GeV) +Au collisions within the statistical model of multifragmentation (SMM). In the second case, the experimental fission probability for excited $^{188}$Os is compared with the calculated one with $T_c$ as a free parameter. It is concluded for both cases that the critical temperature is higher than 16 MeV.

---

[*] Email address: karna@jinr.ru

1. INTRODUCTION

The critical temperature $T_c$ for the liquid-gas phase transition is a crucial characteristic related to the nuclear equation of state. According to [1,2], the nuclear equation of state (EOS) can be presented as follows:

$$p = a\rho + b\rho^2 + c\rho^3, \tag{1}$$

where $a = k_B T$, $b = -k_B T_c/\rho_c$ and $c = 2\, k_B T_c/6\rho_c^2$. The coefficients $b$ and $c$ depend directly on the value of the critical temperature $T_c$ and the critical density $\rho_c$. This EOS is similar to Van der Waals equation suggested in 1875.

There are many calculations of $T_c$ for finite nuclei. In [2-6], it is done by using the Skyrme effective interaction and the thermal Hartree-Fock theory. The values of $T_c$ were found to be in the range 10-20 MeV depending upon the chosen interaction parameters and the details of the model. In Ref. [7,8] the thermostatic properties of nuclei are considered employing the semi-classical nuclear model, based on the Seyler-Blanchard interaction. In [8] critical temperature is estimated to be $T_c = 16.66$ MeV.

As the temperature of a nucleus increases, the surface tension decreases and then *vanishes* at $T_c$. This vanishing defines the critical temperature [9]. For temperatures below the critical one, two distinct nuclear phases coexist - liquid and gas. Beyond $T_c$ there is not a two-phase equilibrium, only the nuclear vapor exists.

Figure 1 shows the different approximations used in the literature for the surface tension coefficient as a function of $T/T_c$. Curve 1 corresponds to the well known equation for $\sigma(T)$:

$$\sigma(T) = \sigma(0)\left[\frac{T_c^2 - T^2}{T_c^2 + T^2}\right]^{5/4} . \qquad (2)$$

This equation was obtained in Ref. [10] devoted to the consideration of thermodynamic properties of a plane interface between liquid and gaseous phases of nuclear matter in equilibrium. This parameterization is successfully used by the statistical model of multifragmentation (SMM) for describing the multi-fragment decay of hot nuclei [11].

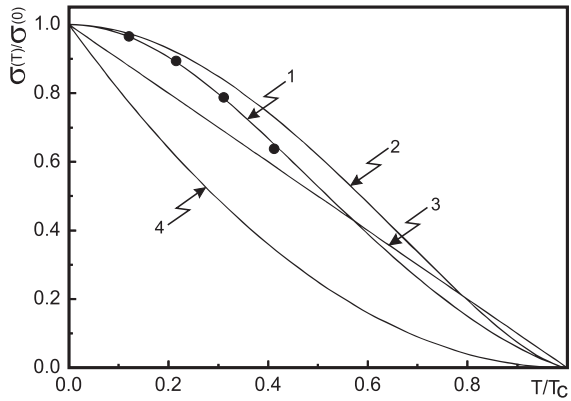

FIG. 1. The calculated coefficient of the surface tension as a function of $T/T_c$: lines 1 and 2 are obtained according to eq. (2) and (3), lines 3, 4 are for linear and quadratic parameterizations of σ (T). The symbols are the calculations from Ref. [3].

Curve 2 was calculated within the framework of the semi-classical model based on the Seyler-Blanchard interaction [7,8]. An analytical expression for σ(*T*) obtained in [8] is

$$\sigma(T) = \sigma(0) \cdot \left(1 + 1.5\frac{T}{T_c}\right) \cdot \left(1 - \frac{T}{T_c}\right)^{1.5} . \qquad (3)$$

Two other parameterizations of σ(*T*) are also presented in Fig. 1: linear, σ(*T*) ~ (1-*T*/*T*c), which is used in the analysis of the multifragmentation data with the Fisher droplet model [12,13], and quadratic, σ(*T*) ~ (1-*T*/*T*c)² [14]. The symbols are taken from the calculations of Ref. [3], which were different from those used in [10]. The agreement of such different calculations supports the reliability of eq. (2).

## 2. CRITICAL TEMPERATURE FROM MULTIFRAGMENTATION DATA

Nuclear multifragmentation is the main source of the experimental information about the critical temperature. In some statistical models the shape of the charge distribution for the intermediate mass fragments (IMF) is sensitive to the ratio $T/T_c$. It was noted in earlier papers that the fragment charge distribution is well described by the power law, $Y(Z) \sim Z^{-\tau}$ [15], as predicted by the Fisher prescription for classical droplets in the *vicinity of the critical point* [16]. In Ref. [15] the critical temperature was estimated to be ~ 5 MeV simply from the fact that the IMF mass distribution is well described by a power-law for the collision of $p$ (80-350 GeV) with Kr and Xe. In [17] experimental data were gathered for different colliding systems to get the temperature dependence of the power law exponent. The temperature was derived from the inverse slope of the fragment energy spectra in the range of the high-energy tail. The minimal value of $\tau$ was obtained at $T$ = 11-12 MeV, which was claimed to be $T_c$. The later data smeared out this minimum. Moreover, it became clear that the "slope temperature" for fragments does not coincide with the thermodynamic one, which is significantly smaller.

A more sophisticated use of Fisher's model has been made in [12]. The model is modified by including the Coulomb energy released when a particle moves from the liquid to the vapor. The data for multi-fragmentation in $\pi$ (8 GeV/c) + Au collisions were analyzed yielding a critical temperature of (6.7±0.2) MeV. The same analysis technique was applied to collisions of Au, La and Kr (at 1.0 GeV per nucleon) with a carbon target [13]. The extracted values of $T_c$ are (7.6±0.2), (7.8±0.2) and (8.1±0.2) MeV respectively. Note that Fisher's prescription is reasonable when the temperature of the system is close to the critical one. In fact, it is not the case. Application of this model may give a spurious value for $T_c$

Actually, the Fisher analysis [12,13] is sensitive to the so called "critical behavior" during fragmentation. In the thermodynamical limit (infinite systems) it really happens in the vicinity of the critical point, where the first-order transition looks like the second order one. However, in finite systems (excited nuclei) the critical behavior can take place in the coexistence region. The shortcomings of Fisher's prescription where already discussed in the literature [21,22].

It should be noted that in some papers the term "critical temperature" is not used in the strict thermodynamic sense given above. In Ref. [18] multifragmentation in Au + Au collisions at 35 MeV per nucleon was analyzed with the so-called Campi plot [19] to prove that the phase transition takes place in the spinodal region. The characteristic temperature for that process was denoted as $T_{crit}$ and found to be equal to (6.0±0.4) MeV. The more appropriate term here is the "breakup temperature". This temperature corresponds to the onset of the fragmentation of the nucleus entering the phase coexistence region. Sometimes the term "limiting temperature" is also used for it. Analysis of the experimental data on the "limiting temperatures" in Ref. [20] resulted in estimation of $T_c$ for the symmetric nuclear matter, which was found to be equal to 16.6±0.86 MeV.

Having in mind the shortcomings of Fisher's prescription we have estimated the nuclear critical temperature within the framework of the statistical multi-fragmentation model, SMM [11]. This model describes well different properties of thermal disintegration of target spectators produced in collisions of relativistic light ions. The yield of intermediate mass fragments, $Y(Z)$, depends on the contribution of the surface free energy to the entropy of a given final state, therefore it is sensitive to the value of the critical temperature. This is well demonstrated in Fig. 2 taken from [23]. Experimental data for $p$(8.1 GeV)+Au collisions are compared with model predictions. The combined INC+SMM model is used with the intranuclear cascade prescription to describe the first stage of the reaction. The comparison of the measured and calculated IMF charge yields provides a way to estimate $T_c$. In Ref. [24] it was done by the analysis of the fragment charge distributions for the $p$(8.1 GeV)+Au reaction with $T_c$ as the only free parameter. It was found that $T_c$= (20 ± 3) MeV.

In the next paper by the FASA collaboration [25] the value $T_c$= (17 ± 2) MeV was obtained from the analysis of the same data using a slightly different separation of the events. In contrast to [24], the analysis was done for the fragments in the range $Z$ = 4 - 11 to exclude the influence of nonequilibrium emission of Li. It was found that the fragment yield depends not only on the critical temperature but also on the breakup volume $V_t$, (Fig.2). The optimal values of $V_t$ and $T_c$ were obtained by the least-square fitting of the data [23]. The results are presented in Fig. 3 showing that $T_c \geq 18$ MeV.

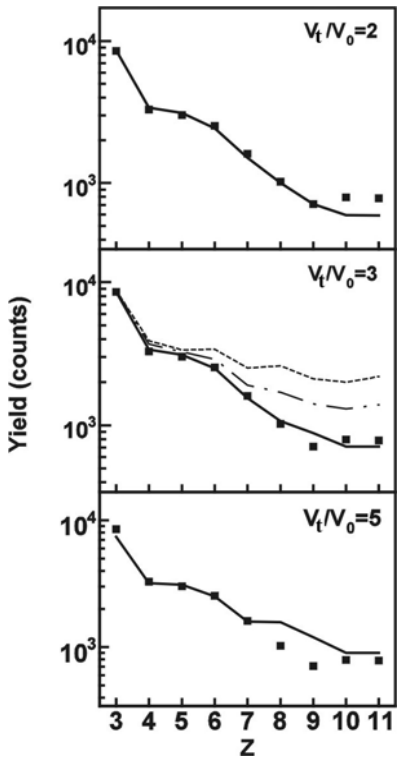

FIG. 2. $p$(8.1GeV)+Au collisions: symbols are for the data; the solid lines are calculated with $T_c$= 18 MeV and break-up volumes indicated; dashed and dot-dashed lies are calculated with $T_c$= 7 and 11MeV.

.

Thus, different experimental estimations of the critical temperature from fragmentation data are very contradictory. This is a reason to look for other observables that are sensitive to the critical temperature for the liquid-gas phase transition. It was suggested in Ref. [26] to analyze the temperature dependence of the fission probability to estimate $T_c$. The following two chapters are based on [27].

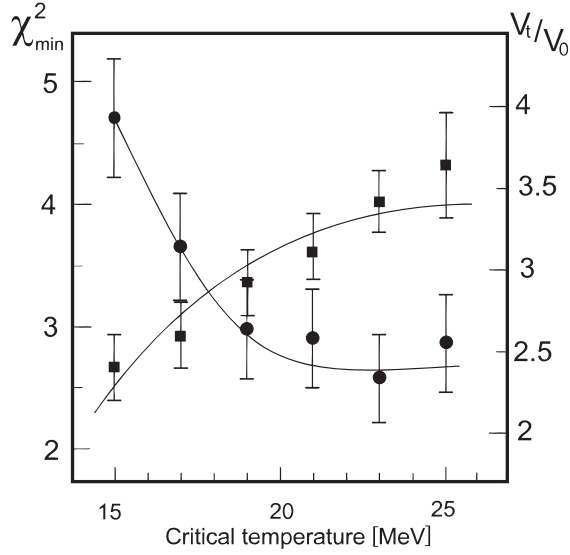

FIG. 3. Results of comparison of measured and calculated IMF charge distributions for $p$(8.1 GeV)+Au collisions with $T_c$ and break-up volume $V_t$ as free parameters. Circles are for the minimal values of $\chi^2$ obtained for a given $T_c$, squares are for the corresponding values of $V_t$.

## 3. TEMPERATURE DEPENDENCE OF FISSION BARRIER

The fissility of heavy nuclei is determined by the ratio of the Coulomb and surface free energies: the larger the ratio, the smaller the fission barrier. As the temperature approaches the critical one from below, the surface tension (and surface energy) gradually decreases, and the fission barrier becomes lower. Thus, the measurement of fission probabilities for different excitation energies allows an estimate of how far the system is from the critical point. Temperature effects in the fission barrier were considered in a number of theoretical studies based on different models (see e.g. [3, 28-33]. The effect is so large for hot nuclei that the barrier vanishes, in fact, at temperatures of 4-6 MeV for the critical temperature $T_c$ in the range 15-18 MeV.

In terms of the standard liquid-drop conventions [34], the fission barrier can be represented as a function of temperature by the following relation:

$$B_f(T,T_s) = E_s(T_s) - E_s^0(T) + E_c(T_s) - E_c^0(T) = E_s^0(T)\left[(B_s - 1) + 2x(T)\cdot(B_c - 1)\right] \qquad (4)$$

Here $B_s$ is the surface (free) energy at the saddle point in units of surface energy $E_s^o(T)$ of a spherical drop; $B_c$ is the Coulomb energy at the saddle deformation in units of Coulomb energy $E_c^o(T)$ of the spherical nucleus; $T_s$ and $T$ are temperatures for the saddle and ground state configurations. For the surface energy and the fissility parameter $x(T)$, one can write [27]:

$$E_s^0(T) = E_S^0(0) \cdot \frac{\sigma(T)}{\sigma(0)} \cdot \left[\frac{\rho(0)}{\rho(T)}\right]^{2/3} \qquad x(T) = \frac{E_c^0(T)}{2E_s^0(T)} = x(0) \cdot \frac{\rho(T)\sigma(0)}{\rho(0)\sigma(T)} \qquad (5)$$

where $\sigma(T)$ and $\rho(T)$ are the surface tension and the mean nuclear density for a given temperature. Equation (4) can be written as:

$$B_f(T, T_s) = B_f(T_s) + \Delta B_f, \qquad \text{where } \Delta B_f = E_s^0(T_s) - E_s^0(T) + E_c^0(T_s) - E_c^0(T) \qquad (6)$$

Here $B_f(T_s)$ is the fission barrier calculated under the assumption that $T_s = T$. In that case the values $B_s$ and $B_c$ are determined by the deformation at the saddle point, which depends on the fissility parameter $x(T)$. These quantities were tabulated by Nix [34] for the full range of the fissility parameter. The value of $\Delta B_f$ is determined by the surface and Coulomb energies of a spherical drop, and can be easily calculated. For $\sigma(T)$ we use approximation (2). In accordance with [34], the expressions for $E_s^0(0)$ and $x(0)$ are taken to be

$$E_s^0(0) = 17{,}9439 \cdot \gamma \cdot A^{2/3} MeV, \qquad x(0) = \frac{Z^2/A}{50.88 \cdot \gamma}, \qquad \gamma = 1 - 1.7826 \cdot \left[\frac{N-Z}{A}\right]^2 \qquad (7)$$

Sauer *et al.* [3] investigated the thermal properties of nuclei by using the Hartree-Fock approximation with the Skyrme force. The equation of state was obtained, which gives the critical temperature $T_c \approx 18$ MeV for finite nuclei. The temperature dependence of the mean nuclear density was found to be $\rho(T) = \rho(0)(1 - \alpha T^2)$, where $\alpha = 1.26 \cdot 10^{-3}$ MeV$^{-2}$. In what follows we shall use this finding for $\rho(T)$.

Figure 4 shows the relative values of the fissility parameter $x(T)$ for $^{188}$Os calculated as a function of the reduced temperature $T/T_c$. This nucleus is chosen because the results can be compared with the well-known experimental [35]. The calculations are performed for the different versions of σ(*T*) mentioned above. A drastic change of nuclear fissility is expected even halfway to the critical point.

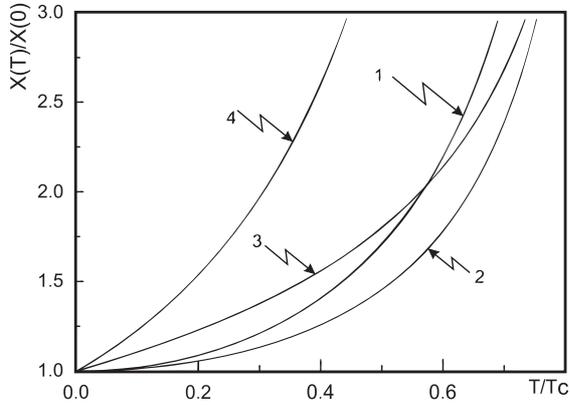

FIG. 4. Relative value of the fissility parameter, calculated for $^{188}$Os as a function of the relative temperature for different parameterizations of surface tension. Notations (lines) are the same as in Fig. 1.

Figure 5 displays the calculated value of the liquid-drop fission barrier for $^{188}$Os as a function of relative temperature. Virtually, the barrier vanishes for $T > 0.4T_c$ if the surface tension is taken according to (2) and (3). For the linear and quadratic approximations of σ (*T*) the reduction of the fission barrier with temperature is much faster.

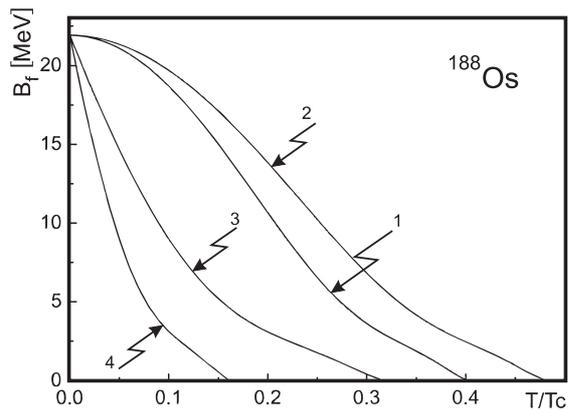

FIG. 5. Temperature dependence of the liquid - drop fission barrier for $^{188}$Os. Notations (lines) are the same as in Fig. 1.

## 4. THE ESTIMATION OF THE FISSION PROBABILITY

In this chapter we analyze the experimental data on the fission probability of $^{188}$Os produced in collisions $^4$He +$^{184}$W [35]. The excitation energy of the compound nucleus created at the highest beam energy is estimated in [30] to be up to 117 MeV. But we considered in the following the data for 40 MeV excitation energy. In this case the compound nucleus cross-section is very close to the reaction cross-section [34], and the excitation energy of $^{188}$Os is directly related to the beam energy. Comparison of the measured and model-calculated fission probabilities provides a way to estimate the critical temperature $T_c$.

Experimentally, the fission probability $W_f$ can be found from the measured fission cross-section $\sigma_f$:

$$W_f = \sigma_f / \sigma_R, \tag{8}$$

where $\sigma_R$ is the total reaction cross-section. The main decay mode of the compound nucleus in $^4$He +$^{184}$W collisions is the sequential emission of neutrons. The mean fission probability during a neutron cascade of $x$ steps can be calculated by the following equation:

$$W_f = 1 - \prod_{i=1}^{x}\left[1 - \frac{\Gamma_f(A_i, Z_i, E_i^*)}{\Gamma_{tot}(A_i, Z_i, E_i^*)}\right], \tag{9}$$

The ratio $\Gamma_f/\Gamma_{tot}$ is the relative fission width for the $i$-th step of the cascade. According to the statistical model [36] the value of $\Gamma_f$ is calculated as

$$\Gamma_f(E_i^*, I_i) = \frac{1}{2\pi \cdot \rho(U_i)} \int_0^{U_i - B_{fi}} \rho_S(U_i - B_{fi} - \varepsilon) d\varepsilon \tag{10}$$

Here $U$ is the thermal part of the excitation energy $E^*$, $\rho(U)$ is the level density, the index $s$ is used for the saddle configuration. It is common to use in Eq.(10) the temperature-dependent fission barrier as was done in Ref. [28-33]. The problem was also considered in [37]. The neutron width is given by the following equation:

$$\Gamma_n(E_i^*, I_i) = \frac{2(2S_n+1)m_n}{\pi^2\hbar^3 \rho_i(U_i)} \int_0^{U_i-B_{ni}} \sigma_n(E_n)\rho_i(U_i - B_{ni} - E_n)E_n dE_n \qquad (11)$$

Here $B_{ni}$, $E_n$, $S_n$ are the binding and kinetic energies and the spin of the neutron, $m_n$ is the neutron mass, $\sigma_n(E_n)$ is the neutron capture cross-section for the inverse reaction. The contribution of charged particle evaporation is at the level of several percent of $\Gamma_{total}$. Nevertheless, it has been taken into account. For the level density $\rho(U)$ the Fermi-gas model is used.

Figure 6 presents the comparison of the data for fissility of $^{188}$Os as a function of the excitation energy [35] with calculations under the assumption that the surface tension is described by Eq. (2). The critical temperature can be obtained from the best fit.

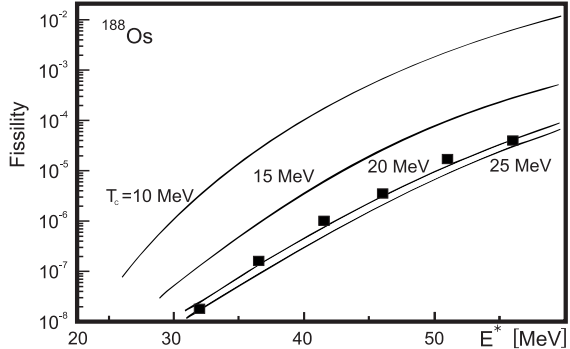

FIG. 6. Fission probability of $^{188}$Os as a function of the excitation energy: dots are data [35], curves are calculated assuming different values of the critical temperature. Surface tension is taken according to Eq. (2).

The result is demonstrated in Fig. 7. Different calculations are presented, which where done using all the parameterizations of the surface tension mentioned above. It seems clear that the linear and quadratic approximations for σ (T) should be excluded as unrealistic ones. The fission probabilities calculated with eq. (2) and (3) fall rather fast with increasing critical temperature. They are crossing the experimental band giving the following values of the critical temperature: $T_c = (19.5 \pm 1.2)$ MeV in the first case, and $T_c = (16.5 \pm 1.0)$ MeV in the second. This agrees with the value of the critical temperature obtained by the FASA collaboration from the multi-fragmentation data. These values only slightly change when the shell effects are taken into account for the last steps of the neutron cascade.

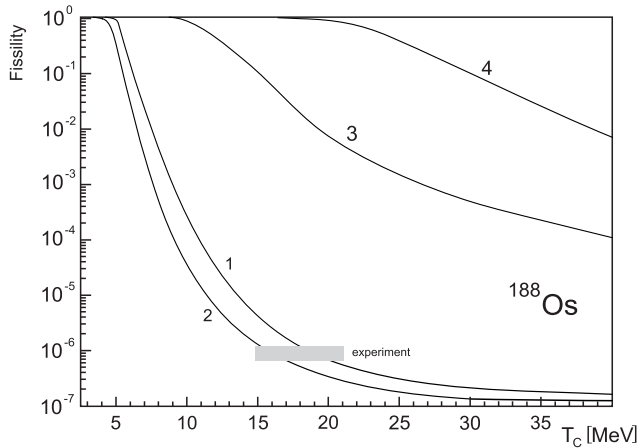

FIG. 7. Fission probability of $^{188}$Os for the excitation energy 40 MeV. The calculated values (lines) are given as a function of the assumed critical temperature. Different parameterizations of surface tension are used (see Fig. 1). The experimental value is shown by the horizontal band.

5. SUMMARY

The critical temperature $T_c$ for the nuclear liquid-gas phase transition is estimated from both the multifragmentation and fission data. In the first case, the critical temperature is obtained by analysis of the yield of intermediate mass fragments in $p$ (8.1 GeV) +Au collisions within the statistical model of multifragmentation. In addition, the experimental fission probability for excited $^{188}$Os is compared with the model-calculated one with $T_c$ as a free parameter. It is concluded for both cases that the critical temperature is higher than 16 MeV. One can say that it is illusive to imagine a nucleus at so high temperature. It is right! But nuclear systems do exist for short time even at higher temperatures, e.g. nuclear fireballs in high-energy nucleus-nucleus collisions. Virtually, the critical temperature is a parameter which determines how fast the surface tension decreases with nucleus heating. The present situation on the "critical temperature market" is rather controversial. It is illustrated by Fig. 8, in which all the data for $T_c$ are collected.

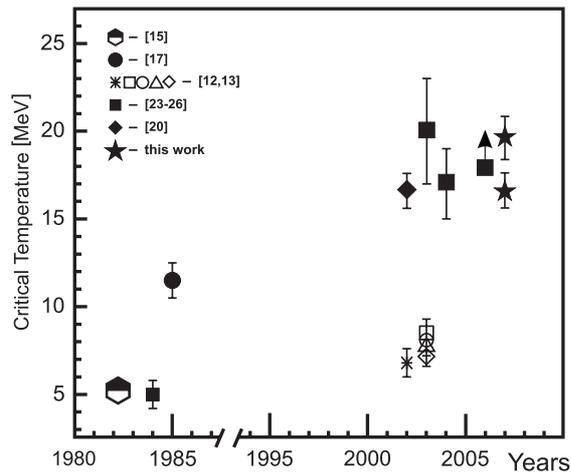

FIG.8. Collection of the experimental data for critical temperature of the nuclear liquid-gas phase transition. All the data are from multifragmentation studies except of "stars" obtained from the fissility of hot nuclei.

The authors are grateful to A. Hrynkiewicz and A.G. Olchevsky for support, to I.N. Mishustin and W. Trautmann for illuminating discussions. The research was supported in part by the Russian Foundation for Basic Research, Grant 06-02-16068 and the Grant of the Polish Plenipotentiary to JINR.

# REFERENCES

[1] J.D. Silva et al., Phys. Rev. **C 69**, 024606 (2004).

[2] A.L. Goodman, J.I. Kapusta and A.Z. Mekjian, Phys. Rev. **C 30,** 851 (1984).

[3] G. Sauer, G. Chandra H. and U. Mosel, Nucl. Phys. **A 264**, 221 (1976).

[4] H. Jaqaman, A.Z. Mekjian and A.Z. Zamick L., Phys. Rev. **C 27,** 2782 (1983).

[5] Zhang Feng Shou, Z. Phys. **A 356,** 163 (1996).

[6] S. Taras et al., Phys. Rev. **C 69**, 014602 (2004).

[7] J. Randrup and E. de Lima Medeiros, Nucl. Phys. **A 529**, 115 (1991).

[8] E. de Lima Medeiros and J. Randrup, Phys. Rev. **C 45**, 372 (1992).

[9] L.D. Landau, E.M. Lifshitz, Statistical Physics, 3d ed., Butterworth-Heineman, Washington, DC, 2000.

[10] D.G. Ravenhall et al., Nucl. Phys. **A 407,** 571 (1983).

[11] A.S. Botvina et al., Yad. Fyz. **42,** 1127 (1985). ; J. P. Bondorf et al., Phys. Rep. **257**, 133 (1995).

[12] J.B. Elliott et al., Phys. Rev. Lett. **88**, 042701 (2002).

[13] J.B. Elliott et al., Phys. Rev. **C 67**, 024609 (2003).

[14] J. Richert and P. Wagner, Phys. Rep. **350**, 1 (2001).

[15] A.S. Hirsch et al., Phys. Rev. **C 29**, 508 (1984).

[16] M.E. Fisher, Physics (N.Y.) **3,** 255 (1967).

[17] A.D. Panagiotou et al., Phys. Rev. **C 31,** 55 (1985).

[18] M. D'Agostino et al., Nucl. Phys. **A 630**, 329 (1999).

[19] H. Campi, Nucl. Phys. **A 495**, 259c (1989).

[20] J. Natowitz et al., Phys. Rev. Lett. **89**, 212701 (2002).

[21] J. Schmelzer, G. Röpke, F.P. Ludwig, Phys. Rev. **C 55**, 1917 (1997).

[22] P.T. Reuter, K.A. Bugaev, Phys. Lett. **B 517,** 233 (2001).

[23] V.A. Karnaukhov et al., Nucl. Phys. **A 780**, 91 (2006).

[24] V.A. Karnaukhov et al., Phys. Rev. **C 67,** 011601(R) (2003).

[25] V.A. Karnaukhov et al., Nucl. Phys. **A 734,** 520 (2004).

[26] V.A. Karnaukhov, Phys. At. Nucl. **60**, 1625 (1997).

[27] E.A. Cherepanov and V.A. Karnaukhov, e-print nucl-th/0703101.

[28] R.W. Hasse and W. Stocker, Phys. Lett. **B 44**, 26 (1973).

[29] A.S. Iljinov, E.A. Cherepanov, S.E. Chigrinov, Z. Phys. **A 287,** 3 (1978).

[30] M. Pi et al., Phys. Rev. **C 26**, 773 (1982).

[31] J. Bartel, P. Quentin, Phys. Lett. **B 152,** 29 (1985).

[32] M. Brack et al., Phys. Rep. **123**, 275 (1985).

[33] F. Garcias et al., Z. Phys. A: At. Nucl. **336**, 31 (1990).

[34] J. R. Nix, Nucl. Phys. **A 130**, 241 (1968).

[35] L.G. Moretto et al., Phys. Lett. **B 38,** 471 (1972).

[36] R. Bass, Nuclear Reactions with Heavy Ions, Springer-Verlag, 1980.

[37] R.J. Charity, Phys. Rev. **C 53**, 512 (1996).